# Formation of relativistic jets by collapsing stars to black holes

Volodymyr Kryvdyk[1]

[1]*Dept. Astronomy, Taras Shevchenko National University of Kyiv, av. Glushkova 2/1, Kyiv 03022, Ukraine*

**Abstract**

Formation of relativistic jets in the magnetosphere of collapsing stars is considered. These jets will be formed in the polar caps of magnetosphere of collapsing star, where the stellar magnetic field increases during the collapse and the charged particles are accelerated. The jets will generate non-thermal radiation. The analysis of dynamics and emission of particles in the stellar magnetosphere under collapse shows that collapsing stars can by powerful sources of relativistic jets.

**Keywords:** collapsing stars, particle acceleration, relativistic jets.

## 1. Introduction

The several models are proposed for relativistic jets from compact objects. Blandford & Znaiek (1977) and Blandford & Paine (1982) assumed that relativistic electron accelerated and jets produced at large distances from holes in magnetized accretion disc by means electromagnetic fields surrounding rotating black holes. In Hanami (1997) magnetic cannonball model the relativistic jets are generated in magnetosphere of compact objects by their collapse to black hole. Koide et al (1998) presented a model of the relativistic jets, in which magnetic fields penetrating the accretion disk around a black and ejection of plasmas. A jet is ejected from a close vicinity to a black hole. The jet has a two-layered shell structure consisting of a fast gas pressure–driven jet in the inner part and a slow magnetically driven jet in the outer part, both of which are collimated by the global poloidal magnetic field penetrating the disk. The former jet is a result of a strong pressure increase due to shock formation in the disk through fast accretion flow. MacFadyen & Woosley (1999) considered the formation of relativistic jets during the evolution of rotating helium stars with mass M > 10 M$_\odot$, in which iron-core collapse does not produce a successful out going shock but instead forms a black hole. In this model simulated deposition of energy in the polar regions results in strong relativistic outflow jets. These jets may be additionally modulated by instabilities. The jets blow aside the accreting material, remain highly focused, and are capable of penetrating the star in about 10 s. After the jet breaks through the surface of the star, highly relativistic relativistic flow can emerge. In paper MacFadyen at al (2001) has studied of the possible production of supernovae and a variety of high-energy transients by black hole formation in massive stars endowed with rotation: the "collapsar model". The energy of the jet and the explosion it produces depend upon the efficiency of MHD processes in extracting accretion energy from the disk. Mizuno et al (2004) performed 2.5-dimensional general relativistic magnetohydrodynamic (MHD) simulations of the gravitational collapse of a magnetized rotating massive star. The simulation results show the formation of a disklike structure and the generation of a jetlike outflow inside the shock wave launched at the core bounce. The jet is accelerated by the magnetic pressure and the centrifugal force and is collimated by the pinching force of the toroidal magnetic field amplified by the rotation and the effect of geometry of the poloidal magnetic field. Magnetohydrodynamic simulations of a rotating massive star collapsing to a black hole are investigated in paper Fujimoto et al. (2006). In this paper authors has perform two-dimensional, axisymmetric, magnetohydrodynamic simulations of the collapse of a rotating star of 40 M$_\odot$ in light of the collapsar model of gamma-ray bursts, and the formation of an accretion disk around a black hole and the jet production near the black hole is investigated. It is found that the jet can be driven by the magnetic fields even if the central core does not rotate as rapidly as previously assumed as long as the outer layers of the star have sufficiently high angular momentum. Meliani at al (2006) studied the steady axisymmetric outflows originating at the hot coronal magnetosphere of a Schwarzschild black hole and surrounding accretion disk in the framework of general relativistic magnetohydrodynamics. In this paper, the authors have investigated of the formation and collimation of relativistic jets. The authors concentrated the efforts on modeling the jet close to its polar axis. In this model it is limited to describing jets possessing a weak rotation velocity compared to the speed of light cylinder. The collimation effects by magnetic and thermal forces and the decollimation effect of centrifugal and electric forces have been studied. It is found that the influence of the electric force and the charge separation in the jet depends on the collimation regime.

In the previous author's papers (Kryvdyk, 1999, 2004) have been investigated the non-thermal emission from the collapsing magnetized stars with dipole magnetic fields and the homogeneous particles distribution in magnetosphere. It is showed that the collapsing stars can be the powerful sources of the non-thermal radiation, which can be observed by means of the modern telescopes.

In this paper we consider the formation of relativistic jets and the radiation by collapse of stars having heterogeneous magnetospheres. The stellar magnetosphere compress during the collapse and its magnetic field increases considerably. A cyclic electric field is produced and the charged particles will accelerate, and the relativistic jets will formed in polar caps of stellar magnetosphere.

## 2. Magnetosphere of collapsing star

We consider the particles dynamics and their radiation from collapsing stars having heterogeneous magnetospheres with the three initial particles heterogeneous distribution in magnetosphere (power-series, relativistic Maxwell and Boltzmann distributions):

$$N_P(E) = K_p r^{-3} E^{-\gamma} \tag{1}$$

$$N_M(E) = K_M r^{-3} E^2 e^{-E/kT} \tag{2}$$

$$N_B(E) = K_B r^{-3} e^{-E/kT} \tag{3}$$

Here $K_P$, $K_M$, $K_B$ are spectral coefficient for power-series, relativistic Maxwell and Boltzmann distributions, $\gamma$ is exponent for power-series distribution, $k$ is Boltzmann constant, $E$ and $T$ - particles energy and temperature, $r$- distance from a centre of star.
The external electromagnetic fields of collapsing stars will change as (Ginzburg and Ozernoy, 1964, Kryvdyk, 1999)

$$B(r,\theta,R) = (1/2)F_0 R r^{-3}(1+3\cos^2\theta)^{1/2},$$
$$E_\varphi = -\frac{1}{cr^2}\frac{\partial \mu}{\partial t}\sin\theta. \tag{4}$$

Where $R(t)$- radius of collapsing stars, $\mu(t) = (1/2)F_0 R(t)$ is a magnetic momentum of the collapsing star, $F_0 = R_0 B_0^2$ – their initial magnetic flux.

In order to investigate the particle dynamics during the collapse we use the method of adiabatic invariant. In this case the particles energy will change as results of the two mechanisms: 1) a betatron acceleration in the variable magnetic field and 2) bremsstrahlung energy losses in this field.
The magnetic moment $\mu(t)$ of collapsing stars change as results of the decrease their radius $R(t)$ under the influence of gravitational field according to the law of free fall

$$dR/dt = (2GM(R_* - 1)/RR_*)^{1/2} \tag{5}$$

$$\frac{d\mu}{dt} = \frac{1}{2}F_0\frac{dR}{dt} = \frac{1}{2}F_0(2GM(R_* - 1)/RR_*)^{1/2} \tag{6}$$

Here $R_* = R_0/R$.

For the heterogeneous distribution the particle dynamics can be investigated using the equation of transitions particle in the regular magnetic fields (Ginzburg and Syrovatskij, 1964)

$$\frac{\partial N}{\partial t} + \frac{\partial}{\partial E}(N\frac{dE}{dt}) + \frac{\partial}{\partial r}(N\frac{\partial r}{\partial t}) = 0 \tag{7}$$

For the new variable $R = R(t)$ this equation becomes

$$\frac{\partial N}{\partial R} + \frac{1}{r^2}\frac{1}{R}f_2(\theta)\frac{\partial}{\partial r}(Nr^3) - \frac{1}{R\sin\theta}\frac{\partial}{\partial \theta}(Nf_3(\theta)) + \frac{\partial}{\partial E}(N\frac{dE}{dR}) = 0, \tag{8}$$

Here $f_2(\theta) = \sin^2\theta(1+3\cos^2\theta)^{-1}$, $f_3(\theta) = (1+3\cos^4\theta)(1+3\cos^2\theta)^{-2}$.

Now we consider temporal changes of particles energy during collapse in the magnetosphere. The particles energy change due to two mechanisms: betatron acceleration in electric field and bremsstrahlung energy losses. For the betatron acceleration in drift approximation we can write (Bakhareva and Tverskoj, 1981)

$$(\frac{dE}{dt})_a = -(1/3)pv\,div\vec{u}, \tag{9}$$

where $p$ is particle impulse, - $\vec{u} = cB^{-2}[\vec{E}\vec{B}]$ the drift velocity.

For the external electromagnetic fields of collapsing stars (4) drift velocity components

$$u_r = \mu^{-1}(\partial\mu/\partial t)r\sin^2\theta(1+3\cos^2\theta)^{-1};$$
$$u_\theta = 2\mu^{-1}(\partial\mu/\partial t)r\sin\theta\cos\theta(1+3\cos^2\theta)^{-1}; \tag{10}$$
$$u_\varphi = 0.$$

Particles velocity field in magnetosphere of collapsing star for Ro/R=100 is given on figure 1. We can see that during collapse the particles outflow with magnetosphere across polar axes, and thus the polar jets are formed by collapse.

From (9) for the field (4) obtained

$$(\frac{dE}{dt})_a = (5/3)\mu^{-1}(\partial\mu/\partial t)pvf(\theta) \tag{11}$$

where $f(\theta) = (3\cos^4\theta + 1.2\cos^2\theta - 1)(1+3\cos^2\theta)^{-2}$

Passing to the new variable $R=R(t)$ according to the law of free fall we can write for energy change

$$(\frac{dE}{dR})_a = (5/3)k_1 f(\theta)\frac{E}{R} \tag{12}$$

For the bremsstrahlung energy losses

$$(\frac{dE}{dR})_s = (e^4/6m^4c^7)F_0^2 g(\theta)R^2 E^2 r^{-6}, \tag{13}$$

where $F_0$ is the total magnetic flux of body, $g(\theta) = (1+3\cos^2\theta)\sin^2\theta$.

The resulting rate of particle energy change in the magnetosphere

$$\frac{dE}{dR} = (\frac{dE}{dR})_a - (\frac{dE}{dR})_s = \frac{5}{3}k_1 f(\theta)\frac{E}{R} - (e^4/6m^4c^7)F_0^2 g(\theta)R^2 E^2 r^{-6} \tag{14}$$

Substitutes Eg. (14) in Eq. (8) we can find the evolution of particles spectrum during collapse. Eq. (8) can not be solved in the general case and so two special cases are considered: (i) when energy losses not influence on the particle spectrum in the magnetosphere and (ii) when the energy losses determine the particle spectrum.
The solution of Eq. (8) in these two cases is given by

$$N^i_P(E,R,r) = K_P r_*^{-3} E_*^{-\gamma} R_*^{-\beta_P}, \qquad (15)$$

$$N^i_M(E,R,r) = K_M r_*^{-3} E_*^2 R_*^{-\beta_M} e^{-E/kT} \qquad (16)$$

$$N^i_B(E,R,r) = K_B r_*^{-3} R_*^{-\beta_B} e^{-E/kT} \qquad (17)$$

$$N^{ii}_P(E,R,r) = K_P r_*^{-3} e^{-\gamma_P} \qquad (18)$$

$$N^{ii}_M(E,R,r) = K_M r_*^{-3} E_*^2 e^{-\gamma_M} \qquad (19)$$

$$N^{ii}_B(E,R,r) = K_B r_*^{-3} R_*^{-\beta_B} e^{-E/kT} \qquad (20)$$

Here $E_* = E/E_o$; $r_* = r_o/r$; $\gamma_P = \gamma(1-\gamma_1)$; $\gamma_M = \gamma_B = (1-\gamma_1)E/kT$; $\beta_P = A_1(\theta)(\gamma-1)$;
$\beta_M = A_1(\theta)(E/kT \ln E_* - 3)$; $\beta_B = A_1(\theta)(E/kT \ln E_* - 1)$; $\gamma_1 = A_2(\theta)F(R,R_*)r^{-6}E$ ;
$A_1(\theta) = (5/3)k_1(3\cos^4\theta + 1.2\cos^2\theta - 1)(1+3\cos^2\theta)^{-2}$;
$A_2(\theta) = (e^4/6m^4c^7)(B_0 R_0)^2 (2GM)^{-1/2}(1+3\cos^2\theta)\sin^2\theta.$;
$F(R,R_*) = \frac{1}{3}R^3(R_*-1)^{1/2} + \frac{5}{12}R^3 R_*(R_*-1)^{1/2} + \frac{5}{8}R^3 R_*^2(R_*-1)^{1/2} + \frac{5}{8}R^3 R_*^3 \arctan(R_*-1)^{1/2}$;
$k_1 = 2$ and $k_1 = 1$ for relativistic and non-relativistic particles respectively.

Eqs. (15) - (17) determine the particle spectrum in the magnetosphere and its evolution during the initial stage of the collapse when the energy losses can be neglected. We will consider this case in this paper. Eqs. (18) - (20) determine the particles spectrum on the final stage of the collapse, when the magnetic field attains an extreme value and the energy losses influence the particle spectrum considerably. This case we will consider in a later paper.

The transformation of the stellar magnetosphere during collapse is should on figures 2. We can see that the initial stellar magnetosphere transform during collapse and the polar jets formed in magnetosphere. A particles density and its energy in polar jets grow during collapse (see figure 3).

**Conclusions**

As follows from obtaining results, the charged particles will accelerate in the magnetosphere of collapsing star to relativistic energy. These particles accumulate in polar caps of the magnetosphere of collapsing stars, and thus relativistic jets will be forms. The density these jets grow during collapse and on the final stage collapse reach the very great values (see figures 2). These jets formed in magnetosphere collapsing stars on the initial stage of collapse before the core collapse.

**References**

Bakhareva, M.F., Tverskoj, B.A. The particles energy variation in the variable interplanetary magnetic field. Geomagn. Aeron. (in Russian), 21, 401-411, 1981.
Blandford, R.D., Znajek, R.L. Electromagnetic extraction of energy from Kerr black holes. MNRAS 179, 433-440, 1977.
Blandford, R. D., Payne, D. G. Hydromagnetic flows from accretion discs and the production of radio jets. MNRAS 199, 883-894. 1982.
Fujimoto, S., Kotake, K., Yamada, S., Hashimoto, M., and Sato, K. Magnetohydrodynamic Simulations of a Rotating Massive Star Collapsing to a Black Hole. Astrophys. J. 644, 1040-1056, 2006.
Ginzburg, V.L., Ozernoy, L.M. About gravitational collapse of magnetic star. Zh.. Exper. i Theor. Fiz.( ZhEThF) 47, 1030-1040, 1964 (In Russian).
Ginzburg, V.L., Syrovatskii, S.I. Origin of cosmic rays. Izd. AN USSR, Moscow, 1963 (In Russian).
Hanami, H. Magnetic Cannonball Model for Gamma-Ray Bursts. Astrophys.J. 491, 687-696, 1997.

Koide, S., Shibata, K., Kudoh, T. General Relativistic Magnetohydrodynamic Simulations of Jets from Black Hole Accretions Disks: Two-Component Jets Driven by Nonsteady Accretion of Magnetized Disks. Astophys.JL 495, L63-66, 1998.

Kryvdyk, V. Electromagnetic radiation from collapsing stars. I. Power- series distribution of particles in magnetospheres. MNRAS 309, 593-598, 1999.

Kryvdyk, V. High- energy emission from presupernova. Adv. Space Res. 2004, 33. 484-486.

MacFadyen, A. I., Woosley, S. E. Collapsars: Gamma-Ray Bursts and Explosions in ``Failed Supernovae'' . Astrophys.J. 524, 262-289, 1999.

MacFadyen, A. I., Woosley, S. E., Heger, A. Supernovae, Jets, and Collapsars. Astrophys. J 550, 410-425, 2001.

Meliani, Z., Sauty, C., Vlahakis, N., Tsinganos, K., Trussoni, E. Nonradial and nonpolytropic astrophysical outflows. VIII. A GRMHD generalization for relativistic jets. A&A, 447, 797-812, 2006.

Mizuno Y., Yamada S., Koide S , and Shibata K. General Relativistic Magnetohydrodynamic Simulations of Collapsars . Astrophys.J. 606, 395-412, 2004.

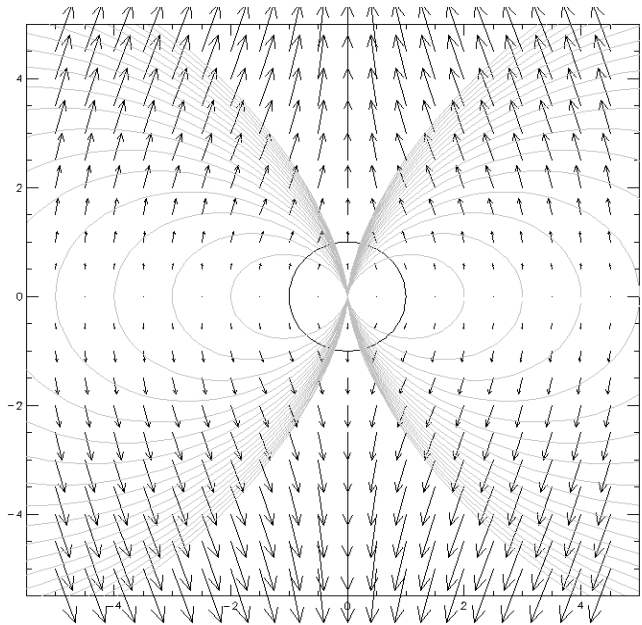

Fifure 1. Particles velocity field in magnetosphere of collapsing star for Ro/R=100.

Fifure 2. Polar jets for Ro/R=1.01, Ro/R=2, Ro/R=5, Ro/R=10. Ro/R=100.

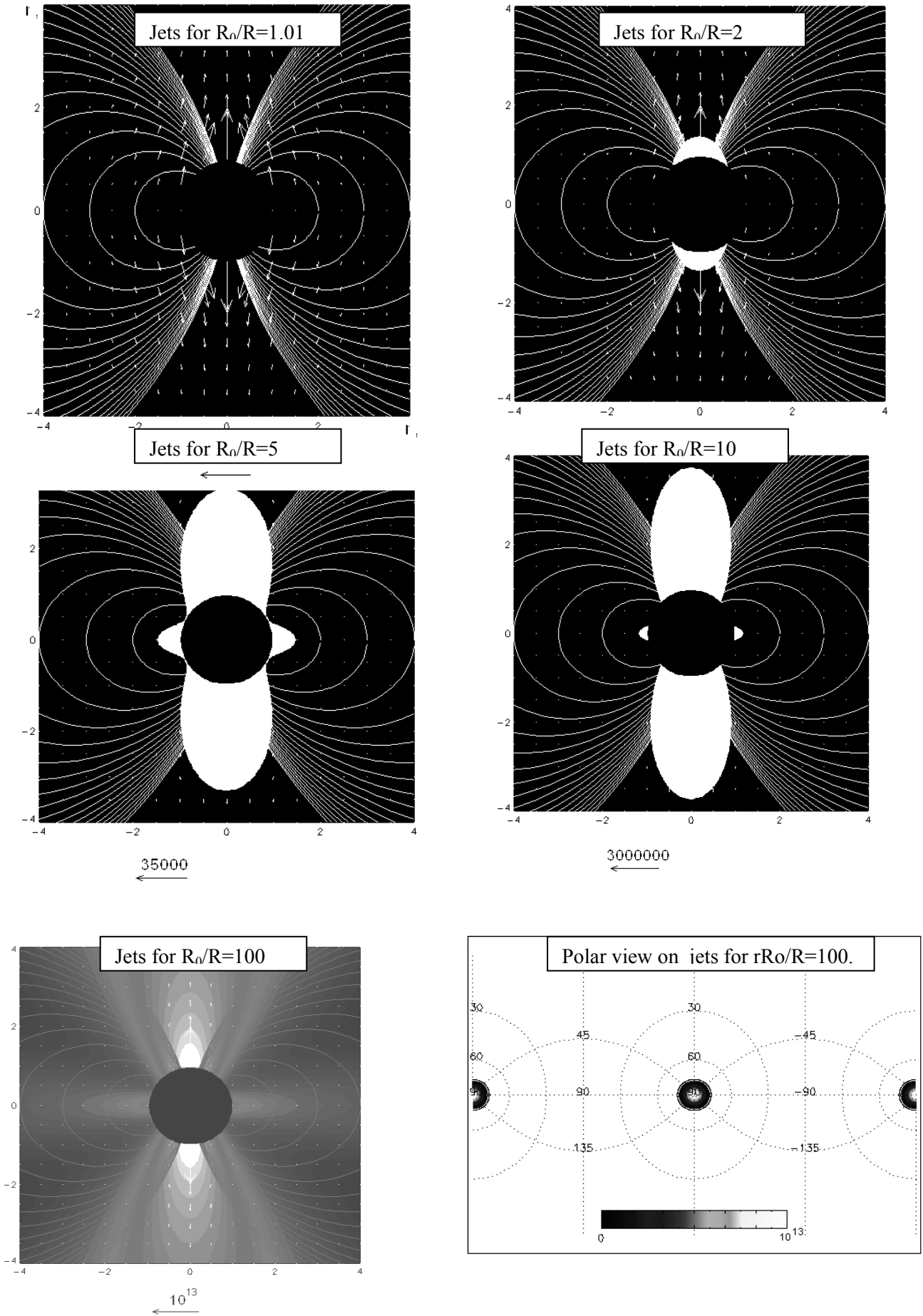

Fifure 3. A increase of the particles density in polar jets during collapse.

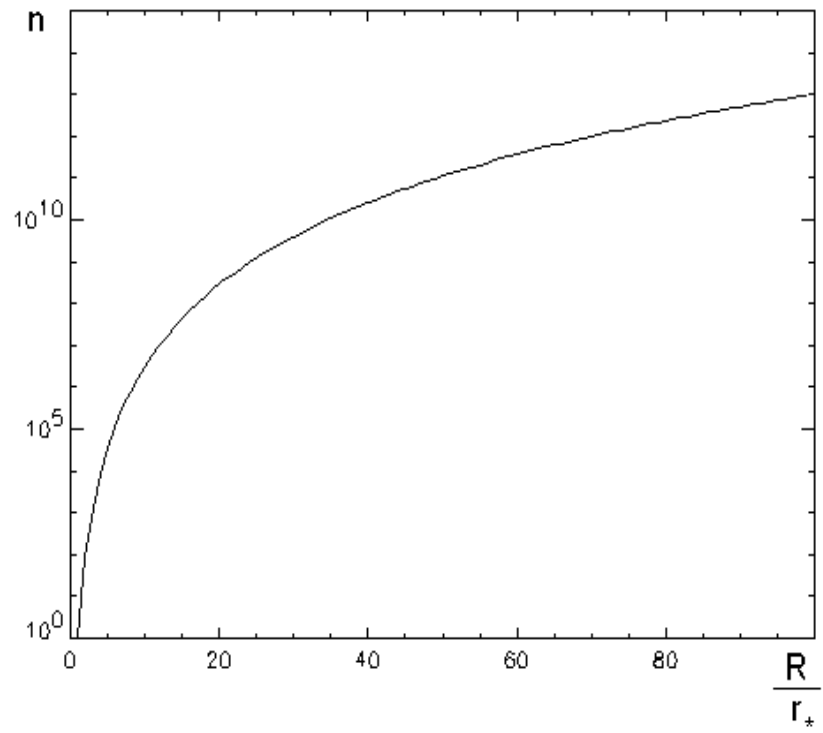